\documentclass[aps,floatfix]{revtex4}
\usepackage{amsmath,amssymb,eucal,graphicx,subfigure,hyperref}
\usepackage{color}
\begin{document}

\title{On the Sensitivity of Protein Data Bank Normal Mode Analysis: An Application to GH10 Xylanases}

\author{Monique M. Tirion}
\email{mmtirion@clarkson.edu}

\affiliation{Physics Department, Clarkson University, Potsdam, New York
13699-5820, USA} 

\begin{abstract}
Protein data bank entries obtain distinct, reproducible flexibility characteristics determined by
normal mode analyses of their three dimensional coordinate files. We study the effectiveness
and sensitivity of this technique by analyzing the results on
one class of glycosidases: family 10 xylanases.  A conserved tryptophan that appears
to affect access to the active site can be in one of two  conformations according to X-ray
crystallographic electron density data.  The two alternate orientations of this active site tryptophan
lead to distinct flexibility spectra, with one orientation thwarting the oscillations seen in the other.
The particular orientation of this sidechain furthermore affects the appearance of the motility
of a distant, C terminal region we term the mallet.  The mallet region is known to separate
members of this family of enzymes into two classes.

\end{abstract}

\maketitle

\section{INTRODUCTION}

\subsection{Protein NMA}

\textcolor{blue}{
Protein atomic coordinates, determined by a variety of techniques, are
deposited in the Protein Data Bank (PDB) \cite{PDB}. The PDB coordinate files permit determination
of numerous physical qualities, 
including charge distribution at a given pH, mass distribution, principal axes of rotation as well as the internal
symmetry axes that characterize each object's flexibility spectrum. Internal symmetry is determined
by normal mode analysis (NMA), a technique first applied to protein structures in the 1980s after the first generations
of (classical) protein force fields matured \cite{go,karplus,levitt85}. 
Go \cite{go} and Levitt \cite{levitt83} independently formulated the eigenvector equations in torsional angle space while
Brooks and Karplus \cite{karplus} studied the vibrational response in Cartesian space.
}

\textcolor{blue}{
As originally formulated, protein NMA requires an energy minimization to bring the PDB coordinates
to a local energy minimum along the force field characterizing the object's energy surface \cite{levitt85}. This necessarily
distorts the PDB coordinates away from the experimentally determined values and precludes the possibility
of a true PDB-NMA.  
Ben-Avraham introduced the use of a (Hookean) potential energy function, one that is a minimum at the outset,
to model inter-monomer interactions in F-actin so as to compute the atomic-based dispersion spectrum of this polymer  \cite{dani}. We then generalized this formulation to  model a protein's
 intra-monomer interactions, effectively introducing the field of PDB-NMA \cite{tirion96}. 
As then formulated, PDB-NMA necessarily retains proper molecular topology:
no distortions of bond lengths or bond angles away from PDB values are allowed due to the use of dihedral degrees of freedom.
That initial formulation examined solely the packing constraints of nonbonded van der Waal
interactions, while current work incorporates restoring potentials associated with each dihedral degree of freedom
as well \cite{na,tirion15}.
}

\textcolor{blue}{
As an alternative to dihedral-based PDB-NMA, various formulations using Cartesian coordinates have been
presented. These are generally referred to as Elastic Network Models (ENM), and include the popular Anisotropic Network Models
(ANM) and Gaussian Network Models (GNM) \cite{bahar,atilgan}.
These alternative approaches to dihedral-based PDB-NMA rely on
structural coarse-graining, such as the use of a single coordinate per residue. While this permits analyses
of vast molecular assemblies such as ribosomes, topological constraints are necessarily sacrificed.
(Even when formulated with full-atomic coordinates, i.e. no coarse-graining, the absence of
bond-length and bond-angle constraints in ENMs will not maintain standard molecular topology.) 
A recent formulation, Torsional Network Model (TNM),  makes use of any suitable potential energy function 
and projects  the Cartesian degrees of freedom onto
dihedral  coordinates in order to preserves topology. Recent reviews summarize the
relative merits of dihedral-based PDB-NMA, ENM and TNM \cite{na,song}.
}

\textcolor{blue}{
While the internal symmetry axes of each PDB entry are of interest in and of themselves, as I try to demonstrate
here, often modal analyses are used to extend or enhance trajectories computed by Molecular Dynamics (MD).
Temporal trajectories, necessarily non-equilibrium pathways, are not well modeled by NMA for two reasons.
NMA is valid only near energy minima and assumes a harmonic or linear response to perturbations.
Efforts to better model the nonlinear aspects of force fields has led to the use of various Principal
Orthogonal Decomposition (POD) algorithms, such as Principal Component Analysis (PCA) and
Singular Value Decomposition (SVD), where eigenvectors are projected out of suitably long-time
MD trajectories \cite{go,wall}.
The degree of overlap between eigenvectors derived from NMA and POD is still
the basis of investigation \cite{pca}.
}

Here we continue to explore the usefulness, range of validity as well as sensitivity of dihedral-based PDB-NMA \cite{tirion15}.
Since no structure-distorting energy minimizations are required, we are able contrast the spectra of structures 
that differ only slightly, such as the two isoforms of the same molecule.
To ascertain the significance of any difference, we will compare the flexibility characteristic of the
same molecule, but solved in a different crystal form and also in the presence and absence of bound ligand.
We also examine the computed flexibility spectrum of an evolutionary distant but structurally similar 
sample of the same enzyme.

We find that PDB-NMA provides detailed, precise, rapid, and reproducible predictions of flexibility
signatures of PDB entries. Our investigation on the flexibility characteristics of one class of enzymes within
the Carbohydrate Active Enzyme (CAZy) database, the glycoside hydrolase family 10 (GH10) xylanases, 
provides a mechanistic rationale for the distribution of experimental
 temperature factors, demonstrates how stability of key residues is maintained in the face of thermal perturbation,
indicates the degree of homology in the flexibility patterns across enzymes with closely similar architecture, and
identifies regions with distinct flexibility patterns.  These analyses suggest that studying a
protein's flexibility characteristics is helpful in order to understand and categorize the unique consequences of
the overall three dimensional conformation of each PDB entry.

\subsection{Protein System Studied}

Xylanases hydrolyze xylan,  wood sugar polysaccharides of the aldopentose xylose.
Unlike planar glucose polysaccharides, xylan adopts a three-fold, left-handed helical
conformation and is often decorated by a variety of branched side chains.
Xylanases are produced by a wide variety of organisms from bacteria to fungi, protozoa and even
gastropods who use xylose as a primary carbon source.  
As each organism targets different sources of plant hemicelluloses and since xylan presents as
complex, branched heteropolysaccharides with structures varying between plant species,
xylanases come with differing sensitivities to structural details of substrate and environmental cues.
Various commercial enterprises, such as the paper
and pulp industries, wine production and brewing, textile and baking industries to
name a few, process the hemicelluloses found in plants. This explains the intense interest
and the large numbers of X-ray crystallographic structures involving xylanases  to 
study their structures and activities under various conditions of temperature, pH, alkalinity etc. \cite{collins}

Xylanases belonging to the GH10 family of glycosidases possess a classic  $(\alpha/\beta )_8$ or TIM barrel fold.
The central barrel, consisting of eight parallel $\beta$-strands, flares out from a narrower 
``stability face''  at the N-terminal ends to
a wider ``catalytic face'' at the C-terminal ends of the $\beta$-strands \cite{hoecker}.
Xylanase GH10 members display their greatest diversity in structure in the positions and loop architecture
at the open, catalytic face, where the substrate binding groove and active site are situated \cite{collins}.  

Family 10 xylanases hydrolyse the internal (1,4) glycosidic bonds linking xylose moieties using two conserved glutamate residues, 
one acting as a general acid/base and the other as a nucleophile.  As one glutamate is situated at the
C-terminal end of $\beta$-strand 4 and the other at the end of $\beta$-strand 7, these xylanases belong to
the 4/7 superfamily of TIM barrel folds (also known as clan GH-A).
Hydrolysis proceeds via a double displacement mechanism that retains the anomeric configuration
of the glycosidic oxygen \cite{rye}. 
This reaction scheme involves formation of a covalently bound, glycosyl-enzyme intermediate, enabling these
xylanases to perform transglycosylation (polymerization) as well as hydrolysis (depolymerization) reactions
\cite{charnock98,moreau}.

The substrates of family 10 xylanases bind in deep grooves to enable proper orientation and distortion of the sugar
moieties to strain the glycosidic bond prior to cleavage.
These binding grooves extend away from the  catalytic glutamate
residues, with ``pockets'' able to accomodate consecutive sugar moieties on both sides of the
cleavage site.  Each ``subsite'' is labeled: ...-2,-1,1,2,..., with negative subsites situated at the
non-reducing, glycone end of the scissile bond, and positive sites situated on the reducing,
aglycone side. The glycosidic bond between the xylose residues bound at subsites -1 and 1 is
cleaved. Family 10 xylanases posses no fewer than  3 subsites, and possibly as many as 6 or 7 \cite{collins}.

Family 10 xylanases possess numerous conserved residues besides the catalytic glutamates, 
including three tryptophans that form an ``aromatic cage'' about subsite -1.
One tryptophan in particular, situated in the loop after $\beta$-strand 8 and present in all family 10 xylanases, 
is unique, with no equivalent aromatic residue  found in any  other 4/7 superfamily of enzymes \cite{leggio}.
Experimental studies demonstrate that mutation or
inactivation by oxidation of this tryptophan disables enzyme activity in xylanases \cite{roberge,privett}.  An early X-ray
crystallographic study of another glycosidase, lysozyme, demonstrated that the oxidation of its
active site tryptophan
 inhibits the binding of substrate and enzyme activity, and results in a rotation of the indole moiety
\cite{blake}.  
Recent X-ray crystallographic studies on family 10 xylanases demonstrate that
the orientation of the active site tryptophan in these enzymes correlates with the presence or absence of ligand in the active site \cite{leggio}.
We were interested to observe whether PDB-NMA 
could detect the  effect of the orientation of this residue
in the vibrational signatures of these enzymes. In fact, we observe a clear difference, with one orientation
of the active site tryptophan leading to a distinct dampening of the oscillatory motion.

\section{TECHNIQUE}

\subsection{Method}

To compute eigenspectra and eigenmodes we maintained the same algorithm and parameterization,  ATMAN, recently introduced for PDB-NMA and described in detail in \cite{tirion15}.
The input data include the atomic coordinates, $\vec{r_i}$, and identities of every entry, $i$, in the PDB listing.
Mainchain amide hydrogens were built in using the program {\it reduce} \cite{reduce}. 
We used standard values for  van der Waals radii $R_{vdWi}$, 
atomic masses, $m_i$, and  van der Waals energy well-depths,  $\epsilon_i$, for each atom type (Table~\ref{Table1}). 

In addition to these atom-specific parameters, two additional inputs are required: 
\textcolor{blue}{
the cut-off distance, $D$ that is added to the sum of the van der Waal radii ($R_{vdW}$) for
nonbonded interactions between atom pairs more than three-bond lengths apart},
as well as a stiffness constraint, $k$, on the sidechain dihedral $\chi$ bonds.
As shown in \cite{tirion15}, large values of $D$ result in vast
numbers of nonbonded interactions (NBI) and lead to eigenspectra that differ substantially from those obtained using
energy-minimized structures. Rather than obtain the characteristic ``soft'' responses in the eigenspectra
frequency range $1-250~cm^{-1}$, a
large value of $D$ in ATMAN yields a very stiff signature extending from
$1-40~cm^{-1}$. To obtain optimal fits of the eigenspectra, therefore,
we maintained a value of $D$ equal to 1.6\AA.
To eliminate instabilities in the diagonalization of the normal mode equations caused by weakly bound
surface sidechains, we used a uniform dihedral stiffness constant of $0.1~kcal/mol$.
Like the atom-specific parameters, both $D$ and $k$ were
maintained to fixed values throughout these analyses (Table~\ref{Table1}).

\begin{table}[h]
\includegraphics[width=0.8\textwidth]{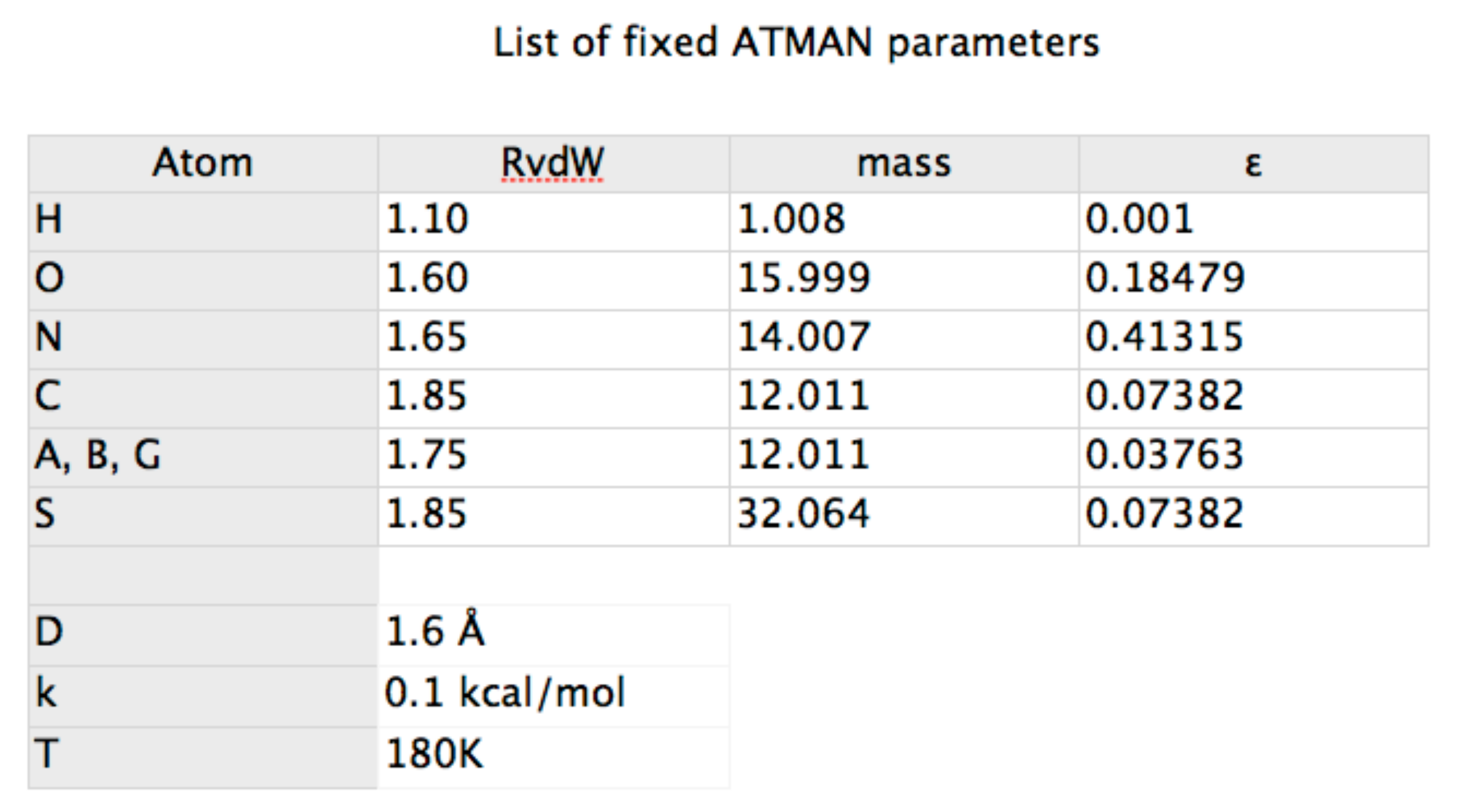}
\caption{H refers to hydrogen atoms, O to oxygen atoms, N to nitrogen atoms, C to tetrahedral carbons
while A, B, and G refer to trigonal carbons and S refers to sulfur atoms. The van der Waals radii are given in
Angstroms, masses are in daltons, and epsilons are in kcal/mol. The atomic data are from \cite{L79,ENCAD}.
The cutoff distance, $D$, and the dihedral bond stiffness constant, $k$, were
as in \cite{tirion15}.  All RMSD were computed at 180K.}
\label{Table1}
\end{table}

In order to reestablish the overall energy scale  of the computed eigenspectra {\it one adjustable} 
parameter, $C$, was necessary to analyze each PDB entry.
Just as proteins across all classes and sizes obtain predictable mass per unit volume measures (density),
proteins also obtain universal eigenfrequency spectra with predictable distributions, especially
of slow modes \cite{universality}. 
We see, for example, both experimentally as well as within the NMA predictions, a distinct peak at
$25~cm^{-1}$ that seems to be due to the inter-packing constraints of secondary elements \cite{tirion15}. 
This peak,
corresponding to vibrations occuring at the $1.3~psec$ time scale, does not correspond to the slowest
motions observed by NMA, which occur on the $1-5~cm^{-1}$ or $30-7~psec$ level, and which
seem to be driven by nonlocal NBIs.  We use the observation
that the number of computed modes with eigenfrequencies under $20~cm^{-1}$ corresponds to 14\% of
all modes, to adjust the overall scale factor, $C$, for the PDB-NMA of each PDB entry.

To visualize the motion described by each eigenfrequency an amplitude of activation, $\alpha_i$ must be selected 
for each mode $i$. Typically this is achieved using
classical conservation of energy considerations: each normal mode  obtains a time-averaged
potential energy of $\frac{1}{2} k_BT$ above the value at NS
($k_B$ is Boltzmann's constant and $T$ is the
absolute temperature).  This gives  $\alpha_i^2 = 2k_BT/w_i^2$ \cite{levitt85}.
We chose a value of $T=180K$ as experiments indicate that only
below this temperature protein motility spectra may be modeled as simple harmonic oscillations \cite{petsko}.

\subsection{PDB entries analyzed}

According to the CAZy database the X-ray crystallographic structures of 37 
GH10 xylanases are currently deposited at the PDB. We report here on the PDB-NMAs
of 5 of these: 1GOKA and 1GOKB, 1GOM, 1I1XA, 1GOO and 1VBR \cite{leggio,natesh,ihsanawati}.
The first 4 entries derive from the thermophilic fungus Thermoascus aurentiacus,
while the final entry derives from the hyperthermophilic anaerobic bacterium Thermotoga maritima.
These PDB entries as GH10 members possess high levels of sequence as well as structural similarities.
As Table~\ref{Table2} shows, the T. aurentiacus xylanases \textcolor{blue} {with 2350 non-hydrogen atoms,} display RMSDs of approximately 0.5\AA~in their
three-dimensional conformations. For 1GOKA and 1GOKB, this RMSD is due to the different
orientation of a small number of sidechains, while the RMSDs between the different crystal structures,
1GOK, 1GOM, 1I1X, and 1GOO, are due to slight structural differences distributed across the entire molecule. 
Even the structure from the hyperthermophilic bacterium, 1VBR, \textcolor{blue} {with 2700 non-hydrogen atoms,}
which obtains a sequence similarity score of 35\% to 1GOKA, obtains a RMSD fit of 1.06\AA~and
scores a structural alignment score of 92\% to 1GOKA \cite{prlic},
emphasizing the structural similarity of these xylanases across species.
Of the 5 PDB entries studied, the 1GOO computations included a ligand, glycerol, in the active site pocket at subsite -1.

\begin{table}[h]
\includegraphics[width=0.8\textwidth]{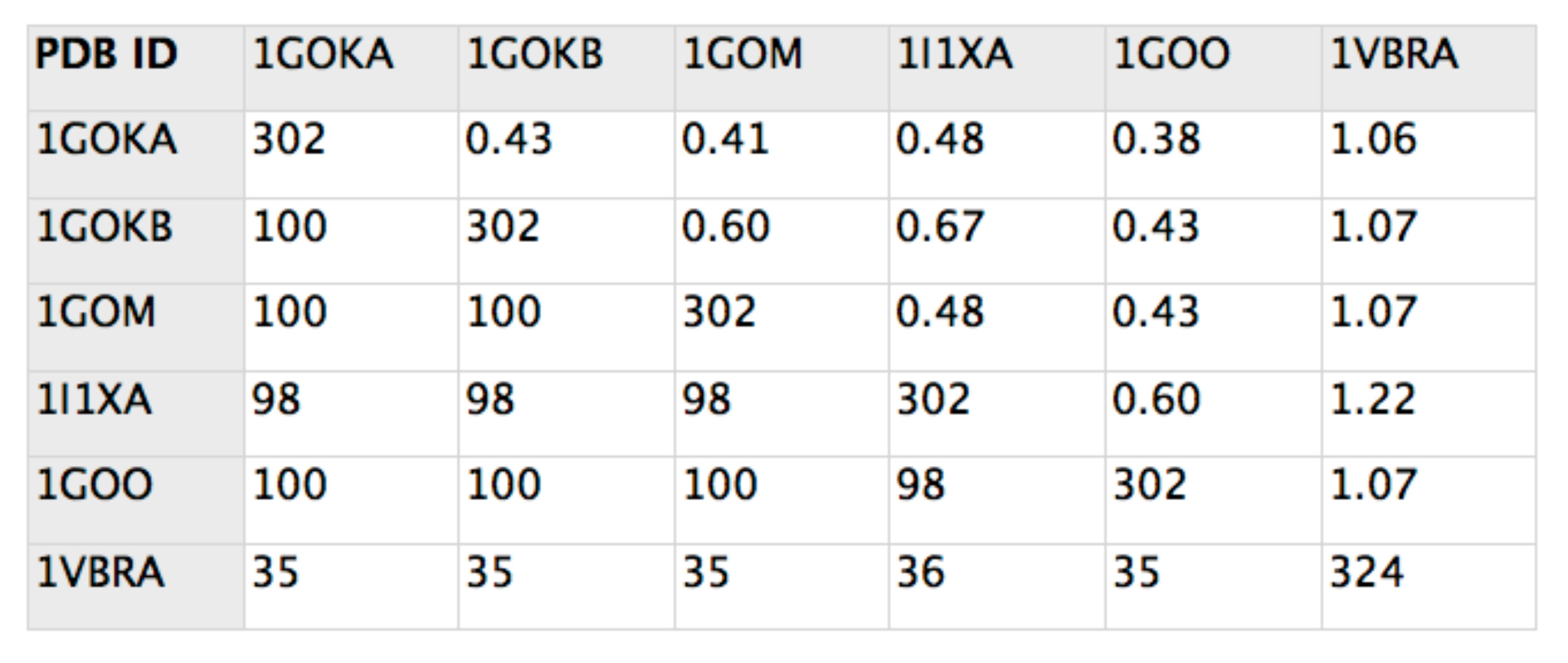}
\caption{Similarities of the 5 PDB entries analyzed.  
RMSD values of all-atom matches recorded above the diagonal.  For
1VBRA RMSD values represent best fits after pairwise sequence matching that eliminates unmatched residues. 
Percentage of primary sequence identities recorded below the diagonal. 
Diagonal entries indicate the number of amino acid residues per entry. RMSD values provided by PyMol. }
\label{Table2}
\end{table}

\begin{figure}[h]
\includegraphics[width=0.5\textwidth]{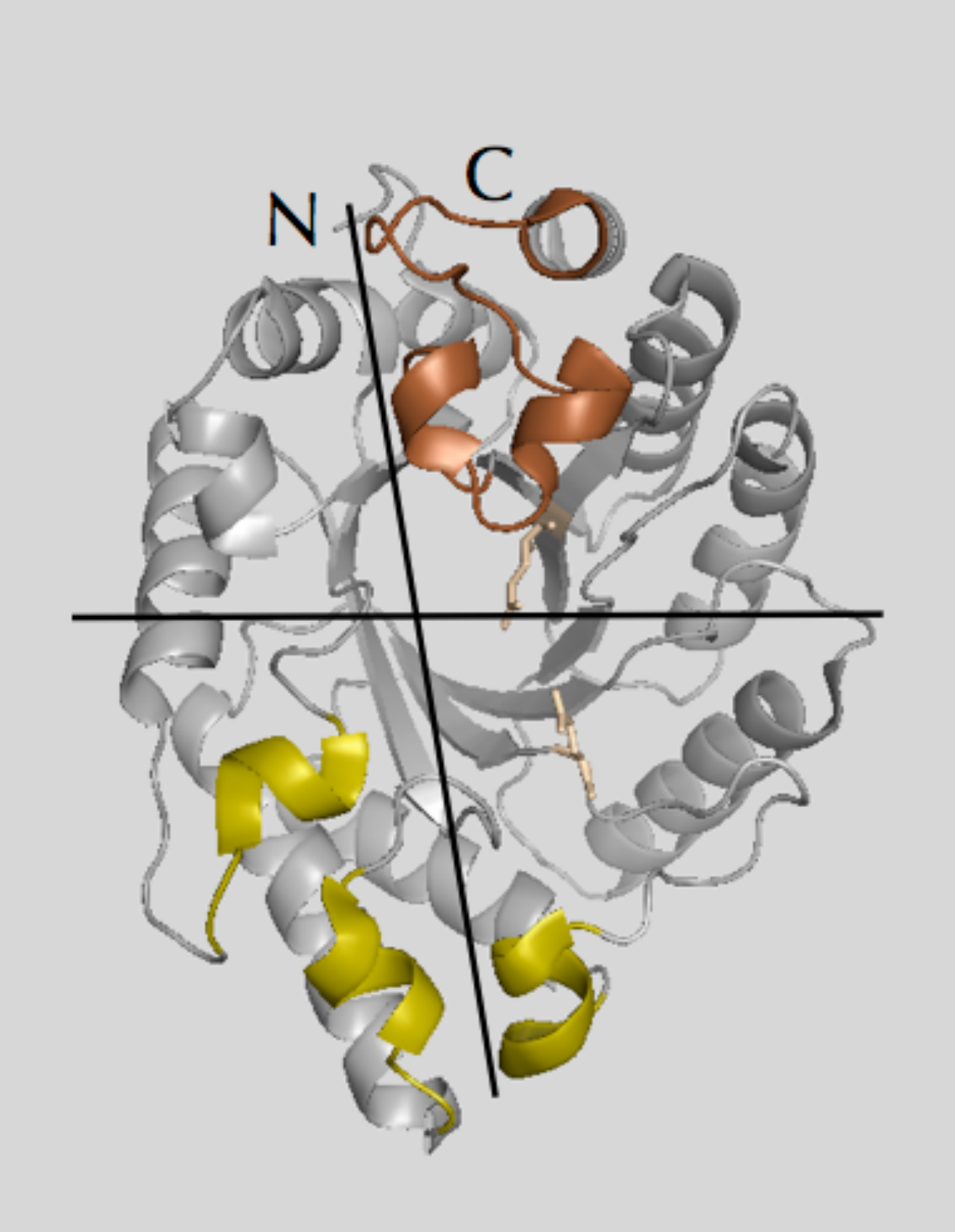}
\caption{Ribbon representation of 1GOK prepared using PyMol. The eight parallel $\beta$-strands of the
TIM barrel open at the catalytic face where the catalytic glutamate sidechains are drawn as sticks.
The binding groove runs roughly along the horizontal axis and separates the upper and lower domain
motion evident in modes 1 and 2.  The nearly vertical axis separates two vertical domains that likewise
oppose eachother in modes 1 and 2.
The C and N terminal regions meet at the top, the ``mallet'' region is highlighted in the upper right quadrant,
while the  three short helices composing``chin'' region are highlighted in the lower two quadrants. Catalytic
Glu 131 falls below the horizontal axis, while Glu 237 is above this axis.
}
\label{Fig1}
\end{figure}

We first examine the flexibility signatures of PDB entry 1GOK in both its A and B isoforms. 
The structure of this 302 amino acid, 2350 nonhydrogen-atom, polymer was determined to 1.14\AA~resolution 
by Lo Leggio and coworkers \cite{leggio}.  Figure~\ref{Fig1} presents its structure in a ribbon representation,
demonstrating the classic $(\alpha/\beta)_8$ fold of this enzyme.  Looking at the structure from
the catalytic face such that the N and C terminal regions meet near the top of the figure, places the
ligand binding groove roughly along the horizontal axis.  This perspective places the catalytic Glu 131 
below and Glu 237 above the axis.
A second axis running nearly perpendicular to the first becomes apparent in animations: this vertical
axis runs from the top, between the noncovalently bound N and C terminal regions, between
$\beta$-strands 1 and 8 and $\beta$-strands 3 and 4 to the bottom-most residue, Asp 100, at the start of
the 3rd $\beta\alpha$ helix. The point of intersection of these two axes in Fig.~\ref{Fig1} marks the location
of subsite -2.

\begin{figure}[h]
\includegraphics[width=0.8\textwidth]{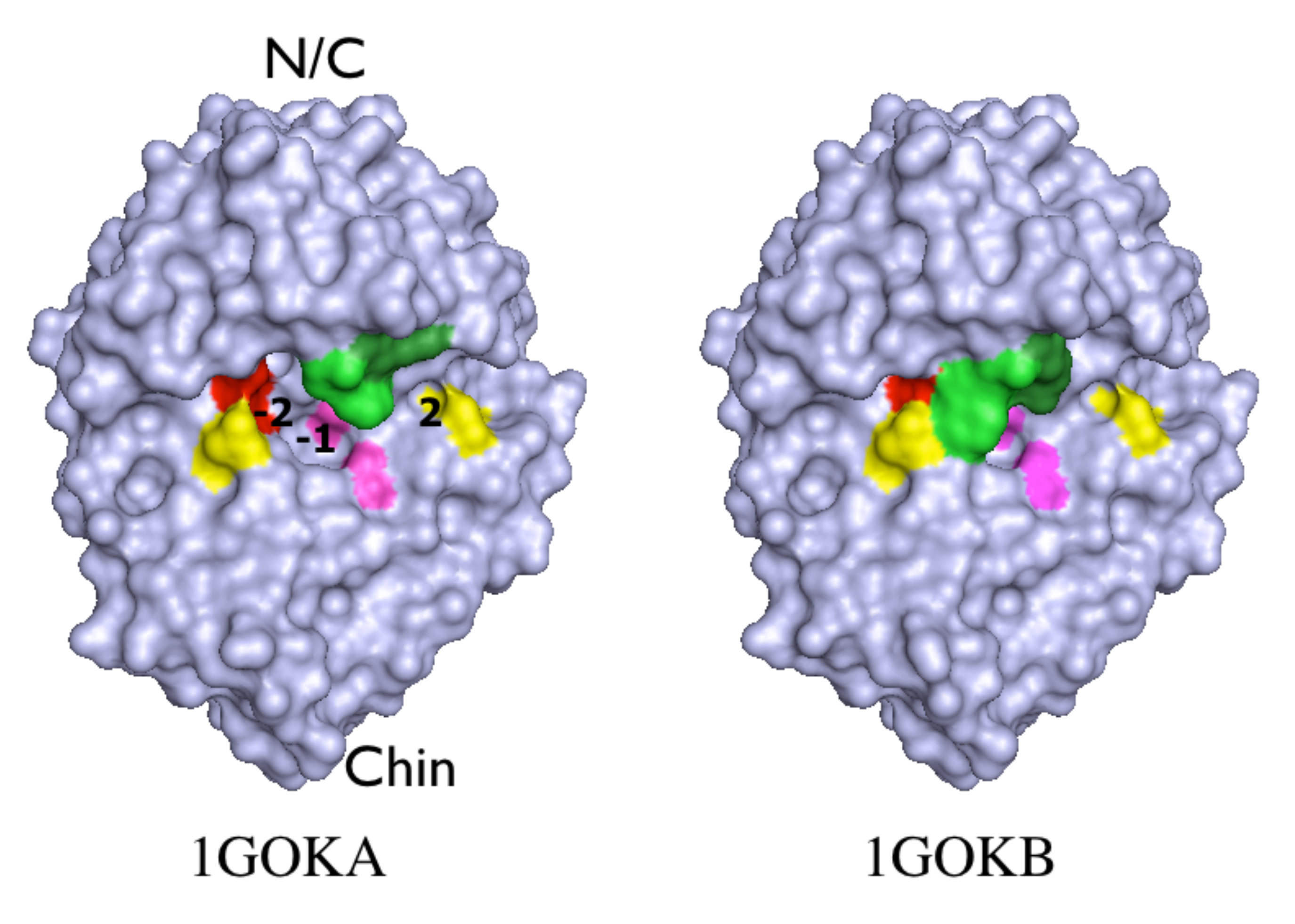}
\caption{Surface representations of 1GOK A (left), with the ``open'' orientation of Phe 275 and Arg 276 in green, and
of 1GOK B, the ``closed'' conformation.  The catalytic glutamates are colored
magenta and Glu 46 in red.  Arg 47 and Arg 175, colored yellow,  are known to be involved in
the primary binding interactions with substrates at subsites -2 and 2.
}
\label{Fig2}
\end{figure}

The electron density of the 1GOK data revealed two approximately equal populations (A and B)
in which 11 residues could adopt one of two conformations \cite{leggio}.   Their all-atom RMSD
is 0.43\AA. Upon inspection,
it is seen that most of this difference is due to three residues adopting distinct conformations (Fig.~\ref{Fig2}).
All three residues, Trp 275 (light green), its neighbor Arg 276 (dark green)  as well as Glu 46 (red), line the binding groove
about subsites -1 and -2, and are strictly conserved among GH10 members.  The remainder of the binding groove residues,
including the catalytic glutamates (magenta) as well as two arginines, 47 and 175 (yellow) known to
promote enzyme-ligand association, are identically situated in the two isoforms.  Indeed,
excluding one sidechain, Trp 275, in calculating the all-atom
RMSD between isoforms A and B reduces the fit from 0.43 to 0.27\AA. If the side chains of all
three residues, Trp 275, Arg 276 and Glu 46 are excluded in the RMSD computation, 
the overall fit further improves to 0.15\AA.

In the 1GOK structures, Trp 275 in particular obtains an interesting
position: at the end of the eighth $\beta\alpha$-loop of the TIM barrel, it projects down and over the subsite -1 hollow.
The image of a lid comes to mind, an image that is re-inforced by the fact that Trp 275B seems to
correspond to a ``shut'' and Trp 275A to an ``open'' orientation of  the lid. 
The apparent lid-like opening and closing that Trp 275 effects is mediated by substantial shifts
in the side chain dihedral values between isoforms A and B:
 $(\chi_1,\chi_2)$= $(70^\circ, -61^\circ)$ in A and $(-75^\circ, 38^\circ)$  in B. The effect is dramatic:
atom CZ2 of Trp 275, for example,
sweeps out an arc of nearly 8\AA~from the A to B conformation. 
To accomodate such substantial local fluctuation, the neighboring Arg 276 side chain likewise shifts
to a substantial degree: its CZ atom shifts by 5\AA~from B to A confomers.  And indeed,
as pointed out by Lo Leggio and coworkers, Trp 275 in its A conformation sterically clashes with
Arg 276 in its B conformation, making their assignments to population A or B unambiguous \cite{leggio}.
Intriguingly, no closed B conformations can be detected for Trp 275 or Arg 276 in structures solved 
with any type of ligand present in the binding groove.

We were interested to observe whether a PDB-NMA could detect differences in eigenspectra and
concomitant eigenmotions due to the small but pertinent shifts of  3 residues out of 302 in
this  high resolution structure. \textcolor{blue}{As we will show}, we find a clear difference in the
mobility spectra of 1GOKA and B: the closed, B conformation dampens
the oscillatory behavior seen in the A conformation. To assess possible relevance of this difference,
we next compute the flexibility spectra of T. aurentiacus xylanases 1GOM and 1I1X that also differ at the level
of RMSD of 0.5\AA, but due to delocalized structural differences.
Again we observe clear and seemingly pertinent trends supporting our observations regarding the B conformation.
We then assess whether the effects observed with  Trp 275 and Arg 276 in their B form is emulated by the
presence of a ligand in subsite -1 by computing the eigenmodes of 1GOO. We find that the presence of
glycerol in the active site cleft does not effect the same dampening as in the B conformations of Trp 275 and Arg 276.
Finally, we will compare the similarities in the flexibilities of fungal (1GOK) versus bacterial (1VBR) xylanases
which further emphasize the possible relevance of observed trends.

\section{RESULTS}

\begin{figure}[h]
\includegraphics[width=0.8\textwidth]{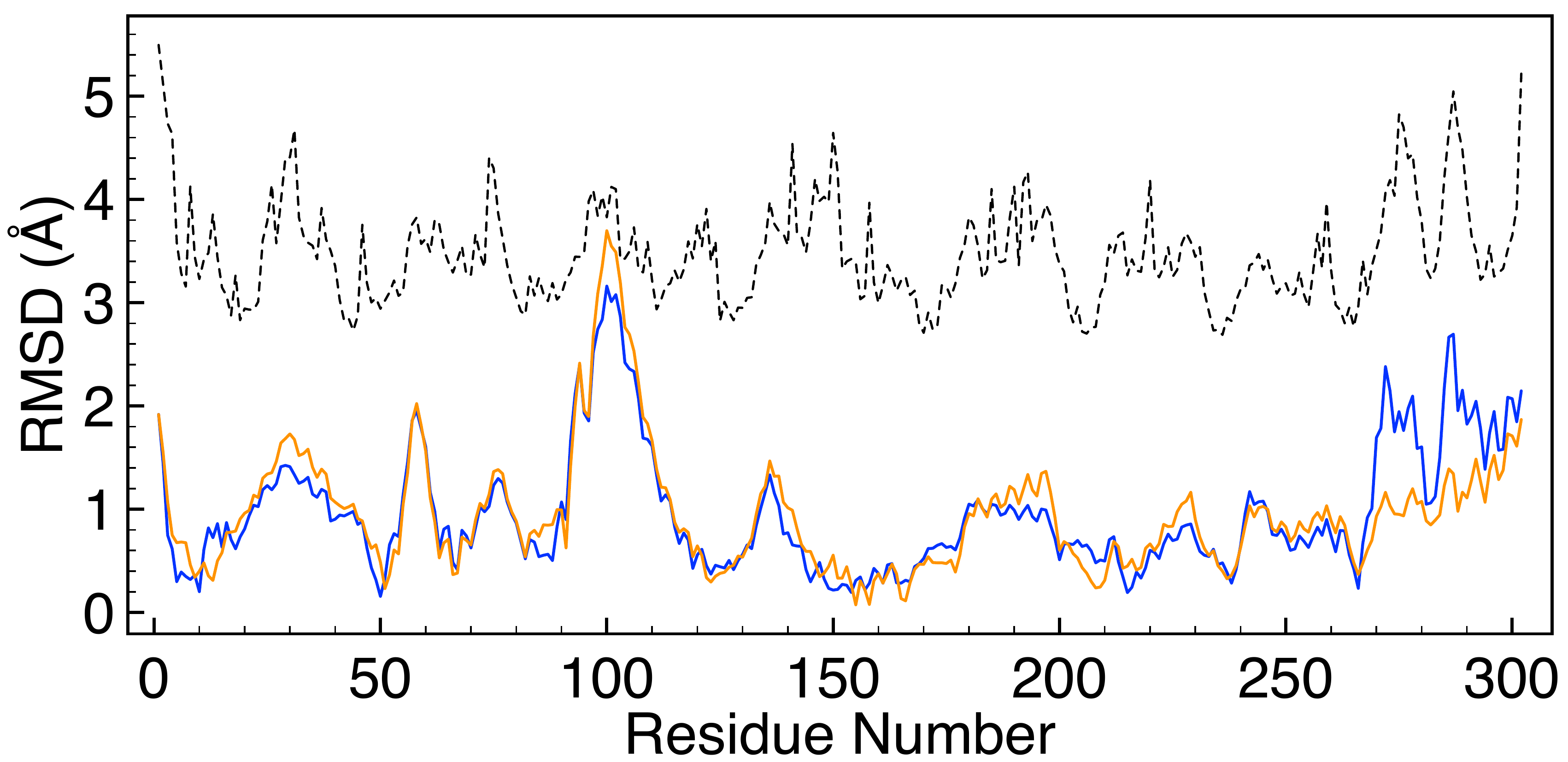}
\caption{RMSD for each $C_\alpha$ due to mode 1 for 1GOKA (blue) and 1GOKB (orange).
The average B factors for each residue, scaled to 10\% the experimental values, are indicated by the dotted line.
}
\label{Fig3}
\end{figure}

Table~\ref{Table3} shows results of the dihedral-based PDB-NMA  on the PDB entries.
1GOKA and 1GOKB obtain 9413 vs. 9390 nonbonded interactions,
with the closed, B, conformation losing 23 NBI  due to loss of alignment of  Trp 275 and Arg 276 
with the 8th and 6th-$\beta\alpha$ loops of the TIM barrel as it swings shut over the subsite -1 cavity.
This 0.25\% decrease in the number of NBI results in an increase of nearly 9\% from
the slowest mode frequency of $2.77~cm^{-1}$ for 1GOKA to $3.02~cm^{-1}$ for 1GOKB. 
Activating the NS according to each eigenvector results  in RMSDs of 0.38\AA~and 0.37\AA~for 1GOKA and B; nearly
identical but slightly bigger for the ``softer'' open A  isomer, despite the smaller $C$ value. 
The surprise, however,
is in the concomitant distribution and largest deviation from NS for all $C_\alpha$.
In Fig.~\ref{Fig3} we plot the RMSD from the NS of each $C_\alpha$ due to thermal activation of
the slowest mode for 1GOKA (blue) and B (orange).  The curves closely overlap, as anticipated for two such
similar structures, with a notable exception  in the C terminal domain.
The curves in Fig.~\ref{Fig3} 
match closely until Trp 267, when the open A confomer suddenly obtains decidedly larger RMS values
than those of the B isomer until Tyr 294 when the RMS values again match until the C terminal residue.  
The transition point, 267, occurs at the C terminal end of the eighth $\beta$-strand (residues 261-266)
and belongs to one of the conserved tryptophan residues lining subsite -1.  
This differently mobile region extending from Trp 267 to Tyr 294 includes the lid residues Trp 275 and Arg 276,
and is highlighted in the upper right quadrant in the ribbon representation of Fig.~\ref{Fig1}. We will refer
to this region as the ``mallet''.
Interestingly, this particular loop structure, the mallet, seems to divide the members of GH10
xylanases into two subsets: molecules with (subset 2) or without (subset 1, like 1GOK) 
an additional $\alpha$-helical stretch immediately after the conserved Trp-275 or its equivalent \cite{leggio}.
Counterintuitively, but
probably due to the loss of mobility of the B confomer at the C terminal region, this confomer, with slightly fewer
NBI and a stiffer $C$, obtains a decidedly larger maximum deviation, at residue 100, than the closed confomer: 3.72\AA~for B
vs. 3.23\AA~for A, a 15\% shift.  Asp 100 and the
short helix-loop preceding it (92-99) as well as the two neighboring short $\alpha$-helical regions (50-58 and
142-149) are highlighted in the lower half of Fig.~\ref{Fig1}, and will be referred to as the ``chin'' region.

\begin{table}[h]
\includegraphics[width=0.5\textwidth]{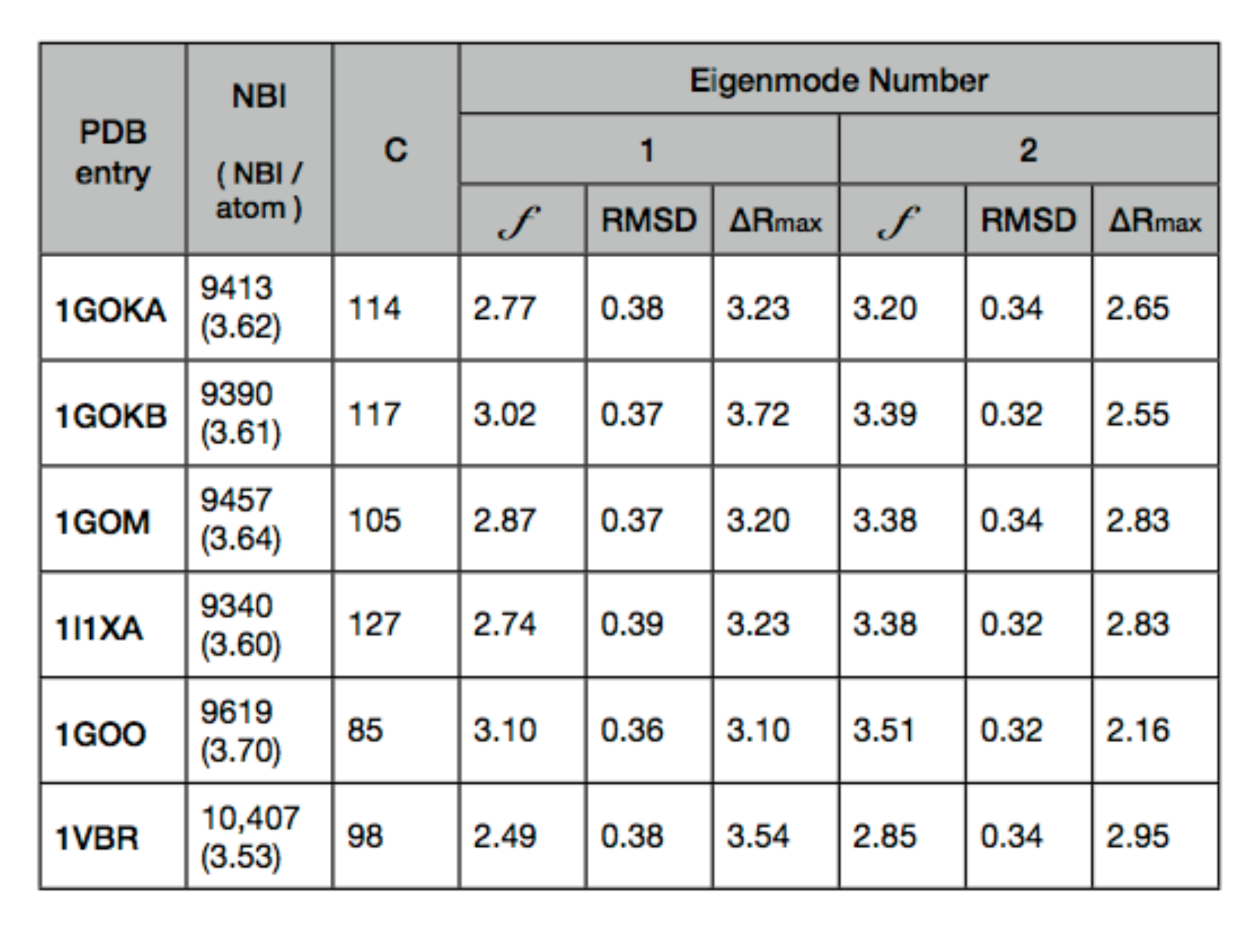}
\caption{Results of dihedral-based PDB-NMA on PDB entries. NBI gives the numbers of nonbonded interactions included for
each computation, NBI/atom gives the average numbers of interaction per atom, and the constant C gives the
overall scale factor.  The frequencies $f$ are given in units of $cm^{-1}$, the RMSD are computed for 180K over
all atoms, and the maximum deviation of any atom from NS is given, in \AA, under $\Delta R_{max}$.
}
\label{Table3}
\end{table}

We studied the mobility patterns via 3d animations to better understand the nature of the computed flexibilities.
The slowest modes of 1GOKA are presented as GIF animations (using LICEcap) in surface
representations (prepared with PyMol)
with the catalytic glutamates in magenta and the subsite -1 tryptophan triad in green in the folders
MODE1 and MODE2 at \cite{gifs} \textcolor{blue}{as well as a pair of stationary images in Fig.~\ref{Fig5}.}
The slowest mode of 1GOK presents as a
``chomping'' motion, Pacman-style,  of the C and N terminal region above the horizontal axis of Fig.~\ref{Fig1} relative to the
region below this axis.  The motion at first glance appears very similar for the 1GOKA and B isoforms, with a
remarkable plasticity surrounding a pronouncedly stable core at the active site. The catalytic
glutamates that remain at a steadfast 5.5\AA~separation, for example, seem enabled by a design scheme that 
deflects innate motility to other portions of the molecule. 
Rather than a motility scheme that compartmentalizes the motions into blocks with tidy divisions between
``fixed'' stability faces and $\beta$-barrels and highly flexible mobility faces, the ``shock-absorption''
is distributed throughout the molecule, with portions adjacent to the stability points deflecting any propensity
to distortions to other regions.  The net effect is remarkable stability for select nonlocal, nonbonded interactions in the face
of ``indiscriminate'' thermal agitation.

Further inspection of the 3d animations also reveals, like the RMS plots, that isoform A obtains a more pronounced 
swinging of the mallet region than isoform B.
Studying the motility pattern, it becomes clear that  several features permit this mallet region 
to obtain such large yet stable RMS deviations.   The residues immediately preceding
the mobile region starting at Trp 267 belong to $\beta$-strand 8 (residues 261-266) which is firmly 
embedded and anchored by the $\beta$-barrel construction.  
The loop immediately preceding this $\beta$-strand, linking   
$\beta$-strand 8 to $\alpha$-helix 7 (residues 244 to 258), is further stabilized by a disulfide bridge linking 
the N terminal residue of $\beta$-strand 8, Cys 261, to the N terminal region of $\alpha$-helix 7 at Cys 255. 
The residues beyond the mobile region ending at Tyr 294 form a C-terminal $\alpha$-helix (residues 291-302) that 
is  anchored between the N-terminal
$\alpha$-helix (residues 5-13)  on one side and the neighboring $\alpha$-helix 7  on the other side, 
and which is itself stabilized, as mentioned, by the disulfide bridge between residues 255 and 261.  
Together,  these features
``ground'' the mobile 267-294 region, maintaining overall structural integrity even as this region itself executes sizeable
RMS excursions.  It is precisely this motion of the mallet which is hampered and dampened in isoform B.

\begin{figure}[h]
\includegraphics[width=0.8\textwidth]{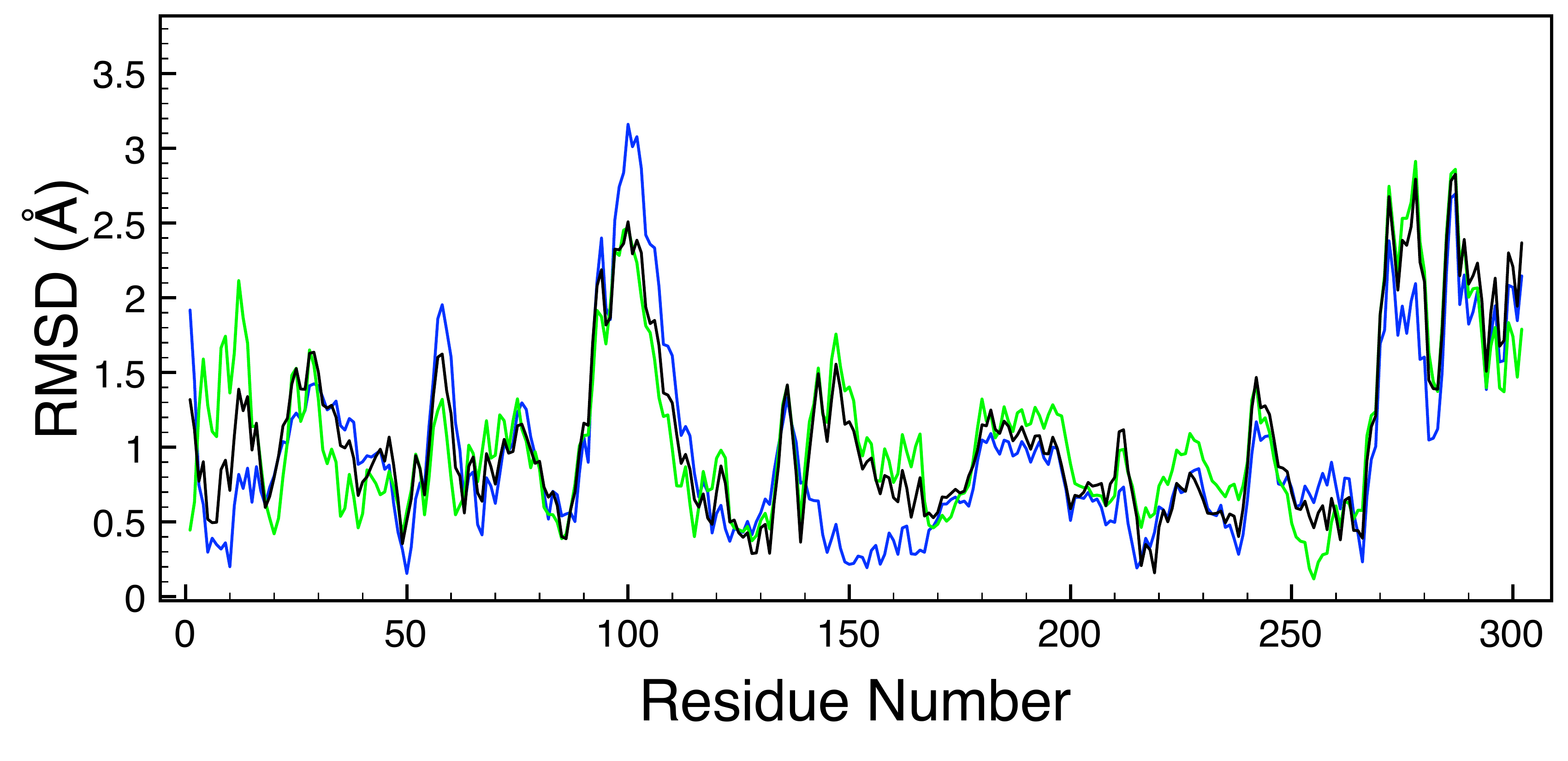}
\caption{RMSD for each $C_\alpha$ due to mode 1 for 1GOKA (blue), 1GOM (green) and 1I1XA (black)
}
\label{Fig4}
\end{figure}

\begin{figure}[h]
\includegraphics[width=0.8\textwidth]{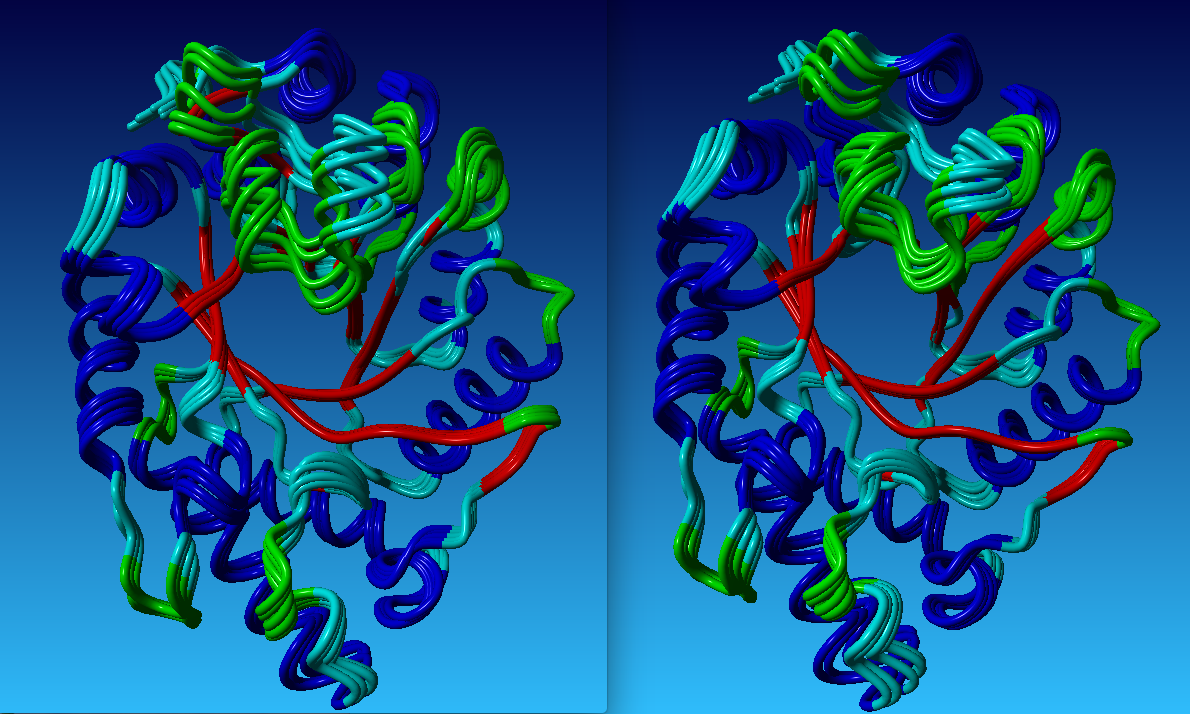}
\caption{Stationary images demonstrating the effects of modes 1 on 1GOKA (left) and 1GOKB (right).
The images, prepared by YASARA, are in the orientation adopted in Figures 1 and 2 and show 5 different
amplitudes of activation of mode 1 on the NS: $\alpha_1\sin(\frac {n\pi}{4})$, with n=0,...,4.  The color scheme is
by secondary structure, with sheets red, helices dark blue, turns green and coils light blue. Regions that
remain stationary have closely overlapping elements, as for the 8 red $\beta$ strands. 
Regions displaying greater motility present separated elements, especially noticeable in the mobile 267-294 region,
the mallet, in the top center as green and light blue of each image. The mallet on the left, of the open 1GOKA, is oscillating
far more than the one on the right, corresponding to the closed 1GOKB.  These significantly differing signatures
are due to the alternate orientations of merely 25 non-hydrogen atoms of Phe-275 and Arg-276, out of 2350 non-hydrogen atoms 
total, in the analysis. This presentation also provides
a sense of the overall harmonicity of the motion, with the central NS structure straddled by the positive and
negative amplitudes of activation of mode 1.
  }
\label{Fig5}
\end{figure}

 The chin region, meanwhile, having the largest moment arm, sweeps out the largest arc and obtains the largest
RMSD in both conformers. Inspecting the disposition of the residues of this region (Fig.~\ref{Fig1}), 
we note that the short $\alpha$-helix (residues 92-97) preceding the highly mobile Asp 100
might move {\it en masse} with the neighboring short $\alpha$-helix at 142-149 at the end of $\beta$-strand 4.
In fact this does not happen in 1GOK as we see a clear tendency of these two regions to pull apart.
The mobile 92-100 chin region rather moves in tandem with the
short $\alpha$-helix-loop above it, residues 50-58.  The reason seems to be due to better stacking interactions
between these two regions, including two indole rings. Trp 51 extends down from the upper, vertical helix
of the chin, while Trp 94 is oriented upwards from the mobile 92-99  helix.
This propensity of the
$\beta$-strands 3 and 4 loops to pull away from each other mirrors that tendency between
the C and N terminal domains, and creates the impression of a dynamic vertical axis, with residues
on one side of the axis tending to pull away from residues on the other side. The point of intersection
of these two axes is interesting, as
residues Glu 46 and Gln 47, important ligand recognition and/or binding residues belonging to subsite -2,
lie on one side of this divide, while Trp 275 and the other residues lining subsite -1 lie on the other
side of this partition. This could result in a possible tension developing along an oligosaccharide spanning
subsites -2 to +2, for example. In the motility pattern of 1GOKB,  there is a decreased chomping
over the horizontal axis, and an increased separation between the vertical domains due to the
alignment of the indole ring of Trp 275 against Glu 46 and Gln 47. The net
effect is a frustration of the chomping mode in 1GOKB and an increased propensity to distort
the shape of subsite -2.

Interestingly, the experimentally observed B-factors of 1GOK, plotted as the dotted curve in Fig.~\ref{Fig3}, seem to 
suggest a large contribution from 1GOKA-type motility, with two peaks in the 267-294 region similar
to those observed in the RMS plot of mode 1 of 1GOKA.  

To test whether these differences in the presentation of mode 1 of 1GOKA and 1GOKB are significant, we
compared their modes to those computed of two other molecules.  One, 1GOM, is of the same protein but
crystallized to a different crystal form (form I rather than II of 1GOK) and solved by Lo Leggio and coworkers
to 1.94\AA ~resolution \cite{leggio}.  1GOM obtains a single conformation, equivalent to 1GOKA, and has an
all-atom RMSD of 0.41\AA~compared to 1GOKA distributed equally over all residues: no single residue mismatches
exist, except for a slight re-orientiation of Trp 275.  The other molecule we examined, 1I1X, is also
of xylanase derived from a T. auriantacus strain, from Indian soil, and has a 297 out of 302 sequence-identity
to 1GOK and 1GOM.  The structure 1I1X was solved by Natesh and coworkers to 1.11\AA~resolution and obtained
two confomers,  A and B \cite{natesh}.  1I1X crystallized to crystal form I, like 1GOM.
The A and B confomers of 1I1X differ in the locations of 23 surface residues:
all active site residues and ligand binding groove residues 
are resolved in one conformation, equivalent to the A confomer of 1GOK. 
We arbitrarily used 1I1XA for the current analysis, which obtained an all-atom RMSD fit to  1GOKA of
0.53\AA, and to 1GOM of 0.48\AA~(Table~\ref{Table2}).  

The number of NBI and C values used for computing the modes of these two proteins are provided in Table~\ref{Table3}.
The frequency of the slowest mode of 1GOM is $2.87~cm^{-1}$ and of 1I1X is $2.74~cm^{-1}$, similar
to the value obtained for 1GOKA ($2.77~cm^{-1}$) whose open Trp 275 structure they resemble.
The RMSD due to thermal activation of each $C_\alpha$ atom in 1GOM  is plotted in Fig.~\ref{Fig4} (green), 
along with that of 1GOKA (blue).
The curves demonstrate very similar mobility patterns, except for mismatches at residues 1-19, 140-168, 
and for the amplitudes for several peaks. Differences in these two curves may be due
to the different crystal packing interactions  
or to the different resolutions of each model. We can ascertain which factor predominates by comparing the
RMSD  plot due to thermal activation of 1GOM to that of the other crystal form I structure, 1I1XA (black).  Here we see that 
the $C_\alpha$ deviations obtained by mode 1 of 1GOM are also obtained by 1I1XA, even though those structures differ
by 0.48\AA. The match includes the 140-168 region, the magnitudes of the peaks, and a tendency for the displacements
of residues 1-19 toward those of 1GOM. In short, these curves demonstrate that in this case crystal packing effects
more strongly correlate with thermal
 RMSD of the slowest mode than with resolution or with the precise details of the atomic
orientations.

To further explore the possibility that Trp 275, by adopting the closed 1GOKB orientation, acts as
a proverbial monkey wrench to block and frustrate the swing of the mallet region, we  wondered whether another agent,
namely the presence of a ligand in subsite -1, in any way mimics the effects of this tryptophan.
Lo Leggio and coworkers solved
the structure of the crystal form II enzyme  cryocooled to 100K \cite{leggio}.  A molecule of the cryoprotectant, glycerol,
a mimic for xylose, was found at subsite -1, while Trp 275 and Asn 276 adopted
the open conformation analagous to 1GOKA.  How does the presence of this  ligand affect
the slowest mode? 
The PDB-NMA of 1GOO, including the additional 6 rigid-body degrees of freedom for the 
glycerol molecule, is unambiguous: the presence of glycerol in subsite -1 does not inhibit the
``swings'' of the mallet region in the way that Trp 275B does.  The RMSD plot of  each $C_\alpha$
due to thermal activation of mode 1 (data not shown) shows that 1GOO obtains a motility spectrum much like
1GOM, with whom it shares an RMS fit of 0.51\AA, including an active mallet region and a
mobile N terminal domain.   The presence of a small ligand in subsite -1 does not eliminate
or frustrate the motion of the mallet region the way that Trp 275B does in 1GOK.

Finally, we examine the flexibility spectrum of a xylanase with an extra $\alpha$-helix 
after the Trp 275 loop, and therefore belonging to subset 2 of xylanases from GH10: 1VBR. 
The hyperthermophilic T. maritima xylanase 10B structure, 1VBR, was solved by Ihsanawati
and coworkers to 1.8\AA~resolution as a dimer \cite{ihsanawati}. We used one copy from the dimer for our analysis.
1VBR obtains 324 residues and is therefore 22 residues longer than the other PDB entries
analyzed. Superposing images of 1VBR and 1GOK reveals added residues in the loops after
$\beta$-strands 8 (the Trp 275-equivalent loop) and 7, and at the end of the C terminal $\alpha$ helix,
lengthening that helix by 5 residues relative to 1GOK.  Aside from these regions, the 
structural overlap between these two structures is excellent: 1GOK achieves a structural match of 98\% to
that of 1VBR,  while  their primary sequences obtain 35\% identity (Table~\ref{Table2}).

The resulting slowest mode bears a striking resemblance, and is essentially identical, to that of 1GOKA, demonstrated by
the stereo images available in the folder $1VBR+1GOK_MODE1$ at \cite{gifs}.  
In this pair of GIF animations,  green traces the mainchain of
1GOKA and cyan traces that of 1VBR.  The sequence is presented in the customary frontal view equivalent
to the orientation in Fig.~\ref{Fig1} as well as from the side after rotation of $90^\circ$ about a vertical axis.

The molecules move in tandem throughout the chain and throughout the sequence, and also obtain
equivalent overall magnitudes of deformation; no additional parameterization was imposed to require
comparable deviations.  Clearly the architecture of the molecular chain
selects for definite flexibility characteristics, irrespective of the particular primary sequence.
These sequences also resolve a puzzle observed with subset 2 Family 10 xylanases: why the Trp 275-
equivalent residues (802 in 1VBR) obtain relatively small temperature factors compared to their
values among the subset 1 members. The active site Trps are not more disordered in subset 1 enzymes
compared to those in subset 2,
they are less mobile than the extended mallet region of these structures.  In 1VBR, the mallet region
once again starts at the Trp located at the C terminal end of $\beta$-strand 8 (residue 794) and extends
to Tyr 826 in the C terminal $\alpha$-helix, but then obtains a further contribution from residues
inserted in the loop sequence after $\beta$-strand 7: Arg 757-Gln 771.  This latter region seem
to ``bulk up'' one end of the mallet while the added residues in the C terminal helix seem to
do the same for the obverse end of the mallet.  It is interesting to note that the stabilizing disulfide
bridge of 1GOK, located toward the stability face of that molecule, has moved in 1VBR. 
A cysteine located at the N-terminal end of $\alpha$-helix 8 (residue 825), is situated to potentially form
a disulfide bridge with Cys 775 in the middle of $\alpha$-helix 7.  In 1VBR however, the distance of
separation of their respective SG atoms precludes the presence of this bond.  In view of the fact that
this molecule derives from a hyperthermophilic organism that thrives in 80C conditions, the absence
of disulfide bridges is surprising.

\section{DISCUSSION}

A number of interesting observations ensue from the study of the PDB-NMA signatures of subset 1 and 2
GH10 xylanases.  Molecules that share a high degree of structural homology display equivalent flexibility
characteristics, as seen in the slowest modes of 1GOKA, 1GOM, 1GOO, and 1VBR. 
The slowest mode pertains to a chomping motion and the second slowest mode pertains to a
grinding motion across the ligand binding groove \cite{gifs,tirion15}.
Plots of  RMS deviations from NS due to thermal activation as well as 3d animations of eigenvectors
demonstrate that select residues remain largely immobile while other portions of the molecule display
sizeable excursions about their NS.  The OE2 atoms of the catalytic residues Glu 237 and Glu 131, for example,
have a distance of separation of 5.57\AA~ in the NS.  Thermal activation of mode 1 in the current protocol 
suggests this distance may increase by 0.04\AA~ or diminish by 0.11\AA. In contrast, the neighboring
subsite -1 residues Trp 275 and Trp 87 behave very differently. CE3 of Trp 275 is a distance of 8.45\AA~
from CH3 of Trp 87 in the NS, and approach as close as 7.29\AA~ and separate as far as 12.26\AA, a net
difference of nearly 5\AA~compared to a net difference of 0.15\AA~ for the neighboring catalytic glutamates. It needs
to be stressed that this criterion is not in any manner built into or required by the computation. This
feature is an expression of packing constraints that uniquely define the internal symmetry
axes of the molecule.  The surprising immobility of key, nonbonded interactions permitted by the
overall molecular design and resultant intra-molecular packing, may indeed have driven the selection of
 protein design.

While select NBI remain relatively immobile during thermal agitation,  other regions display concerted movement, 
such as the ligand binding groove and the C terminal mallet region.  The substrate binding groove especially
experiences considerable realignment.  The slowest mode demonstrates an enhanced likelihood
for the binding groove to be more and less solvent exposed as the N/C terminal, upper domain and lower
 domain close across the binding groove with a characteristic period of around 12 psec.  In addition, the
slowest mode  reveals an interesting region that we refer to as the mallet, a region 
after $\beta$-strand 8 that obtains RMS shifts larger than other parts of
the C/N domain. This particularly mobile region happens to segregate the GH10 members  into two subsets
depending on the nature of their loop design.  Surprisingly, 
the mallet's motility is suppressed by the realignment of the indole ring of a {\it single} tryptophan, a
realignment observed in the A and B confomers of the same crystal structure, 1GOK.  
What might be the consequences of such a mobile mallet region?
A more thorough examination of the known GH10 structures might reveal connections between the
size and mobility of this region and the nature of ligand bound, such as the degrees of polymerization and
decoration for example.

And how to interpret the different Trp 275 orientations in light of these analyses?
Experimental results demonstrate xylanase inactivation by oxidation of the active site tryptophan \cite{woolridge}.
Furthermore, as pointed out, the X-ray crystallographic  structure of the TIM-barrel GH enzyme, lysozyme, revealed
that the oxidation and inactivation of its active site tryptophan resulted in a concomitant rotation of the indole moiety.
This suggests a possibility that the 1GOKB structure represents an oxidized, inactive form of the enzyme.
However, a different interpretation might be provided by the realignment of the active site 
tryptophan seen in lipases. Lipase activity is
dependent on the orientation of a tryptophan residue in a surface loop, called the lid, being  situated  to either
 sterically hinder substrate access to the active site or to permit
access to the active site  \cite{bourbon}. A comparison of modes between these two enzymes might 
therefore prove of interest.

\section{CONCLUSION}

PDB-NMA provides a direct, rapid and reproducible means to sensitively probe the flexibility
signatures of PDB entries.  The realignment of one or two residues out of 300 is sufficient to
create distinct mobility spectra, and likely have direct biochemical consequences in terms of
the activity or inactivity of enzymes, for example.  On the other hand, PDB entries with a
large degree of structural homology display very similar flexibility characteristics, even
in the absence of a high degree of primary sequence homology.  

We have examined primarily the slowest mode of several PDB entries. 
The shape of the slowest mode is principally due to the collective effect of  thousands of nonbonded interactions and is therefore least sensitive to
details in the potential energy formulation used to construct the eigenvalue equation.  Higher frequencies
probe greater details in the atomic potential and therefore require greater care in their formulation and computation.
We here showed that
the slowest eigenmode provides valuable insights into the unique flexibility characteristics of a particular molecular
design, the TIM barrel fold in family 10 xylanases.
We demonstated that select NBI display little to no deformation under thermal agitation while
other regions obtain large deformations. Observed flexibility patterns highlight regions that move
{\it en masse} and bring into focus  features removed from the active site that may nonetheless affect  enzyme function.
Efforts to correlate flexibility patterns and their timescales to enzyme functionality may enhance
understanding of protein design.

\acknowledgements

This work is supported in part by the M. Hildred Blewett Fellowship of the American Physical
Society, \url{http://www.aps.org}.

%

\end{document}